\documentclass[conference]{IEEEtran}
\IEEEoverridecommandlockouts
\usepackage{cite}
\usepackage{amsmath,amssymb,amsfonts}
\usepackage{algorithmic}
\usepackage{graphicx}
\usepackage{textcomp}
\usepackage{xcolor}

\usepackage{multirow}
\usepackage{url}
\usepackage{hyperref}

\usepackage{amssymb} 
\newboolean{showcomments}
\setboolean{showcomments}{true} 

\ifthenelse{\boolean{showcomments}}
{
	\newcommand{\nb}[3]{
		{\colorbox{#2}{\bfseries\sffamily\scriptsize\textcolor{white}{#1}}}
		{\textcolor{#2}{$\blacktriangleright$\textsf\small{#3}$\blacktriangleleft$}}}
	 
}{
	\newcommand{\nb}[3]{}
} 

\definecolor{bgcolor}{RGB}{230, 240, 255}

\definecolor{gray50}{gray}{.5}
\definecolor{gray40}{gray}{.6}
\definecolor{gray30}{gray}{.7}
\definecolor{gray20}{gray}{.8}
\definecolor{gray10}{gray}{.9}
\definecolor{gray05}{gray}{.95}

\newlength\Linewidth
\def\findlength{\setlength\Linewidth\linewidth
	\addtolength\Linewidth{-4\fboxrule}
	\addtolength\Linewidth{-3\fboxsep}
}

\newenvironment{rqbox}{\par\begingroup
	\setlength{\fboxsep}{5pt}\findlength
	\setbox0=\vbox\bgroup\noindent
	\hsize=0.95\linewidth
	\begin{minipage}{0.95\linewidth}\normalsize}
	{\end{minipage}\egroup
	\textcolor{gray20}{\fboxsep1.5pt\fbox
		{\fboxsep5pt\colorbox{gray05}{\normalcolor\box0}}}
	\endgroup\par\noindent
	\normalcolor\ignorespacesafterend}

\usepackage{marginnote}

\usepackage[
    textcolor=blue,
    linecolor=blue,
    bordercolor=blue,
    backgroundcolor=white,
]{todonotes}

\usepackage{soul}
\setstcolor{blue}

\newboolean{highlight}
\setboolean{highlight}{true}

\ifthenelse{\boolean{highlight}}
  {\newcommand{\rref}[1]{\todo{\textcolor{blue}{#1}}}
   
   \newcommand\remove[1]{\textcolor{lightgray}{#1}}
   }
  {\newcommand{\rref}[1]{}
   
   \newcommand\remove[1]{}}
   
\def\BibTeX{{\rm B\kern-.05em{\sc i\kern-.025em b}\kern-.08em
    T\kern-.1667em\lower.7ex\hbox{E}\kern-.125emX}}
\begin{document}

\title{Perspectives, Needs and Challenges for Sustainable Software Engineering Teams: A FinServ Case Study }

\author{\IEEEauthorblockN{Satwik Ghanta}
\IEEEauthorblockA{\textit{School of Computing Science} \\
\textit{University of Glasgow}\\
United Kingdom \\
y.ghanta.1@research.gla.ac.uk}
\and
\IEEEauthorblockN{Peggy Gregory}
\IEEEauthorblockA{\textit{School of Computing Science} \\
\textit{University of Glasgow}\\
United Kingdom \\
Peggy.Gregory@glasgow.ac.uk}
\and
\IEEEauthorblockN{ G\"{u}l \c{C}al{\i}kl{\i}}
\IEEEauthorblockA{\textit{School of Computing Science} \\
\textit{University of Glasgow}\\
United Kingdom \\
HandanGul.Calikli@glasgow.ac.uk}
}

\maketitle

\begin{abstract}
\textbf{Background:} Sustainable Software Engineering (SSE) is slowly becoming an industry need for reasons including reputation enhancement, improved profits and more efficient practices. However, SSE has many definitions, and this is a challenge for organisations trying to build a common and broadly agreed understanding of the term. Although much research effort has gone into identifying general SSE practices, there is a gap in understanding the sustainability needs of specific organisational contexts, such as financial services, which are highly data-driven, operate under strict regulatory requirements, and handle millions of transactions day to day.
\textbf{Aim:} To address this gap, our research focuses on a financial services company (FinServCo) that invited us to investigate perceptions of sustainability in their IT function: how it could be put into practice, who is responsible for it, and what the challenges are.
\textbf{Method:} We conducted an exploratory qualitative case study using interviews and a focus group with six higher management employees and 16 software engineers comprising various experience levels from junior developers to team leaders.
\textbf{Results:} Our study found a clear divergence in how sustainability is perceived between organisational levels. Higher management emphasised technical and economic sustainability, focusing on cloud migration and business continuity through data availability. In contrast, developers highlighted human-centric concerns such as workload management and stress reduction. Scepticism toward organisational initiatives was also evident, with some developers viewing them as a PR strategy. Key challenges included internal knowledge gaps, cultural resistance, legacy system constraints, and a lack of client-driven demand. Many participants expressed a preference for a dedicated sustainability team, drawing analogies to internal structures for security governance.
\textbf{Conclusions:} SSE is shaped by organisational roles, cultural dynamics, and sector-specific constraints. The disconnect between organisational goals and individual developer needs highlights the importance of context-sensitive, co-designed interventions. We contribute novel insights into how SSE is negotiated in a financial services setting, exposing trade-offs, scepticism, and competing incentives.
\textbf{Study material} containing codebook and the interview questions file: \url{https://doi.org/10.5281/zenodo.15185346}
\end{abstract}


\begin{figure*}[htbp]
\centerline{\includegraphics[width=14 cm]{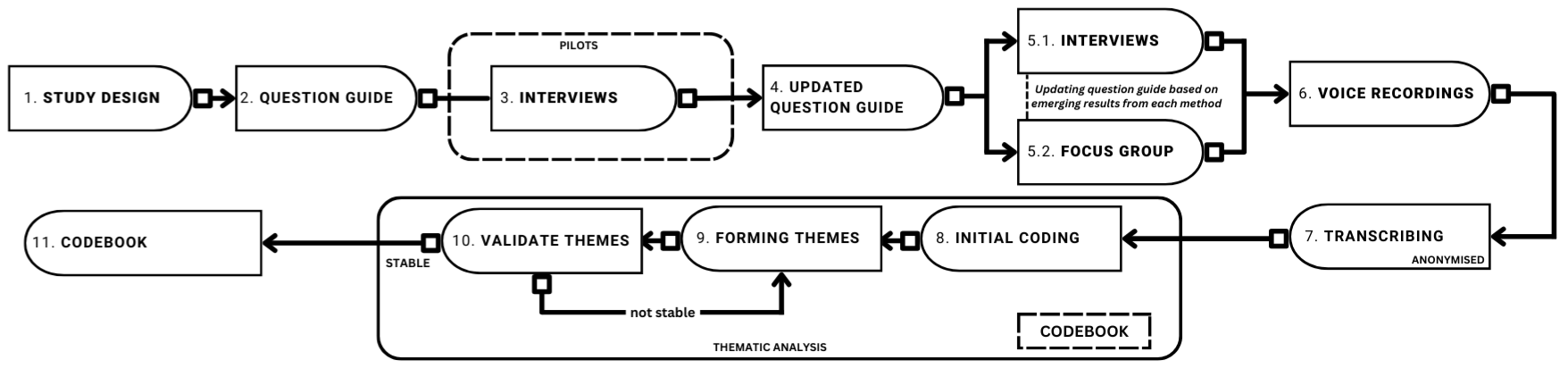}}
\caption{Study design}
\label{fig:research-design}
\end{figure*}
\section{Introduction}

The 1986 World Commission on Environment and Development (WCED) United Nations (UN) report, which is also known as the Brundtland report, defined sustainable development as ``development that meets the needs of the present without compromising the ability of future generations to meet their own needs'' \cite{b11}. The definition in the Brundtland report was accompanied by two key concepts which were later widely ignored in the literature: \textit{`needs'} and \textit{`limitations}.' The concept of \textit{`needs'} proposed that the highest priority should be given to the needs of the world's poor, and the concept of \textit{`limitations'} suggests that the environment's ability to meet people's needs is limited by the current state of technology and social organisation \cite{b11}. These two concepts indicate how \textit{ideal sustainability} (the idea that every design and every system could be one hundred percent sustainable for all people across all dimensions) is not possible. The Brundtland report also explains why priorities and trade-offs are sometimes needed to achieve sustainability.

The term \textit{``sustainability''} (also referred to as \textit{sustainable development} and \textit{sustainable design} by many researchers) has been defined in many different ways over time as it is a broad concept \cite{b12}. The UN defines \textit{``sustainability"} using three dimensions: \textit{Environmental} (e.g., raw resources, climate change, food production, water, pollution, waste), \textit{Social} (e.g., social equity, justice, employment, democracy), and \textit{Economic} (e.g., wealth creation, prosperity, proﬁtability, capital investment, income) \cite{b21, b12}. More recently, Becker et al. \cite{b12} added two new dimensions: \textit{Individual} (e.g., mental and physical well-being, education, self-respect, skills, mobility), and \textit{Technical} (e.g., maintenance, innovation, obsolescence, data integrity within information systems). The authors proposed a manifesto for sustainability design for software and cross-disciplinary systems.

Our primary objective for this exploratory qualitative case study was to examine perceptions of SSE within a financial services organisation, FinServCo that wants to integrate SSE practices into its workflow. Software Engineering (SE) plays a crucial role in modern businesses. IT systems, which have been critical for banks and other financial companies for some time, are becoming increasingly sophisticated due to the introduction of innovative technologies such as AI and increasing competitiveness in the financial environment. Software systems that enhance data security, fast data retrieval, accurate but fair AI, and reduce data duplication are critical to financial organisations, such as FinServCo. Since lack of a common view of sustainability and variance among individuals' perceptions can prevent integrating SSE into an organisations's software engineering practices, we sought to discover the perspectives of both: \textbf{(i)} \textit{managers}, who are policy makers in a company, and \textbf{(ii)} \textit{software developers}, who put these policies into action. A case study approach is particularly appropriate for this purpose, as it can provide an in-depth understanding of a phenomenon (SSE) within an organisation, facilitates the collection of rich, contextualised insights, and captures the nuances of both higher management and developer perspectives within the specific organisational environment \cite{case_study_runeson}. Previous research on SSE practices and frameworks has taken a generalised approach across industries. However, integrating SSE into an organisation's SE practices requires investigating the perspectives,  needs and challenges specific to that organisation's context. Therefore, to the best of our knowledge, we conducted the first case study undertaken in a specific company to investigate sustainability perspectives, needs, and challenges. More specifically, through this study we want to address, \textbf{how employees of a Large-Scale Multi-National Financial Services Organisation (FinServCo) interpret SSE and perceive its implications for integration into working practices.}

\subsection{Data Availability}
This study adheres to Open Science principles. The interview protocol, information leaflet and codebook are publicly available\cite{b35}. However, interview transcripts cannot be shared due to ethical restrictions imposed by FinServCo to ensure data protection and participant confidentiality. Also, a detailed list of follow-up questions is not shared, as they were asked dynamically during semi-structured interviews and are different for each participant.

\section{Related Work}

In this section, we cover the related work in the SE literature that investigated software professionals' perspectives of sustainability, and industry needs and the solutions proposed to integrate sustainability into Software Development Lifecycle (SDLC). 

\noindent{\textbf{SE Practitioners' Mindset towards SSE.}}
Various studies have investigated the perspectives of software engineers. Betz et al. \cite{b6} and Heldal et al.~\cite{b4} investigated developers from different industries and roles; whereas Groher and Weinreich~\cite{b7} conducted interviews with technical leads in different industries, and Souza et al.~\cite{b9} interviewed research software engineers within one university. Along with these, Chitchyan et al.~\cite{b8} investigated the perspectives of requirement engineers across various industries.

Betz et al.~\cite{b6} and Heldal et al.~\cite{b4} found that practitioners and organisations have a poor understanding of sustainability. Heldal et al. also found a lack of proper knowledge sharing between teams which impacted sustainability.  According to the findings of Groher and Weinreich \cite{b7}, most technical leads view SSE from a technical perspective as persistent systems ignoring other dimensions.  Similarly, Souza et al. \cite{b9} found that research software engineers viewed sustainability as the ability for the project to run for longer, but, in contrast to Groher and Weinreich \cite{b7}, their participants also identified sustainability as reusability of software artefacts by other teams. Souza et al. \cite{b9} classified characteristics of sustainable software into: Intrinsic (e.g., documented, easy to use, testable) and Extrinsic (e.g., openly available, well-resourced, actively maintained) sustainability. 
The findings of Heldal et al. indicate that many organisations look at sustainability from an economic perspective, while consumers and/or clients are more aware of the environmental and social perspectives.  
\smallskip

\noindent{\textbf{Industry Needs.}}
Heldal et al. \cite{b4} focused on industry needs and the sustainability education and skills required by SE practitioners in industry to fulfil the sustainability goals of their organisations. They interviewed 28 participants from different organisations across different sectors, either one-to-one or in focus groups. Key findings indicate that while many organisations were interested in incorporating sustainability into their practices, there was a poor understanding of the concepts. This study also reported that many organisations struggle with short-term profits versus long-term sustainability trade-offs, mainly because of a lack of proper awareness or metrics to understand sustainability. They also noted the need for in-house training courses and collaboration with universities to fill the sustainability knowledge gap. 

\smallskip

\noindent{\textbf{Software Development Life Cycle (SDLC).}}
Chitchyan et al. \cite{b8} provided intervention strategies for handling sustainability blockers during Requirements Engineering (RE) and mentioned that developers' stress and high workloads could hinder sustainability changes. Levy et al. \cite{b5} also investigated  the RE phase, similar to Chitchyan et al. \cite{b8}, but focussed on healthcare systems. In contrast, the study of Karita et al. \cite{b33} covers all phases of SDLC. On the other hand, Ibrahim et al. \cite{b32} focussed on software development processes. Levy et al. interviewed RE developers across various industries and healthcare professionals, whereas the other two studies involved participants from different industries.

Levy et al. \cite{b5} developed a Sustainable Health Requirements Engineering (SusHeRE) Framework which identifies four RE goals for sustainability and health: \textbf{(i)} \textit{RE Techniques} that can be used to elicit requirements while keeping human needs and sustainability in mind, \textbf{(ii)} \textit{Multidisciplinary Expertise} domains from which knowledge can be identified and used to create inclusive sustainability solutions, \textbf{(iii)} \textit{Education Agenda} containing RE topics to be taught to equip practitioners with the skills needed to design sustainable solutions, \textbf{(iv)} Public and Social Ecology regarding the skills RE experts need to communicate with and influence policy makers to embed sustainability into their decisions.

Similarly, Ibrahim et al. \cite{b32} proposed a Green Software Process Model which provides green factors (benchmarks) for the software development process. This aims to promote sustainability and minimise negative environmental impacts, including excessive energy and resource consumption, often overlooked in current practices. The findings highlight the need for a more comprehensive integration of sustainability into software processes and further empirical studies to refine and validate the model's effectiveness in real-world settings.

On the other hand, Karita et al. \cite{b33} found that sustainability concerns are present in all SDLC phases. They emphasise the importance of incorporating sustainability early in the SDLC, through stakeholder engagement and quality requirements gathering, to support sustainable development. They also note that while sustainability is gaining recognition as a software quality attribute, there is still a lack of consensus on how it can be integrated into traditional software development processes.

\subsection{Research Gap}

All the studies discussed above took a generalised approach to data gathering, recruiting participants from different organisations and industries, except Souza et al. \cite{b9} who studied research software developers in Universities and Levy et al. \cite{b5} who conducted a workshop with RE and healthcare professionals. Yet, none of the mentioned studies aimed to understand the needs and challenges of a specific organisation to facilitate the integration of SSE into its workflow. In addition, there is a gap in understanding whether and how top-level managers' (policy makers) perspectives differ from bottom-level employees' (policy followers). Understanding the differences between the SSE perspectives of policy makers and policy followers is crucial since, without a shared understanding, making SSE integral to developers' daily practices becomes a challenge. Therefore, in this study, we address the mentioned gaps in the literature by presenting the case study of SSE perspectives in one organisation within the Financial Services industry.

\section{Research Methodology}

This section presents our research questions and provides information on the study design, including information on interviews, demographics of the participants, data collection, and analysis. The study material containing the codebook and interviews are available online~\cite{b35}.

\subsection{Research Questions}
To explore \textbf{how employees of a Large-Scale Multi-National Financial Services Organisation interpret SSE and perceive its implications for integration into working practices}, we conducted an exploratory qualitative case study at FinServCo. Through this study, we address the following five sub-questions:
\smallskip

\begin{center}
    \begin{rqbox}	    
       \textbf{RQ$_1$.} \emph{What does sustainability mean for higher management in FinServCo?}
    \end{rqbox}
\end{center} 

This question aims to explore what higher management (policy makers) understand about sustainability, and what they envision for their organisation.

\begin{center}
    \begin{rqbox}	    
       \textbf{RQ$_2$.} \emph{How do software practitioners interpret sustainability in FinServCo?}
    \end{rqbox}
\end{center} 

This question looks at the perspectives of the software developers (policy practitioners) and team leads (policy enforcers) on sustainability, and their perception of the organisational view on sustainability.

\begin{center}
    \begin{rqbox}	    
       \textbf{RQ$_3$.} \emph{What are the perceived needs for incorporating sustainability into software engineering teams?}
    \end{rqbox}
\end{center} 

This question aims to understand what aspects are prioritised by different groups of participants.

\begin{center}
    \begin{rqbox}	    
       \textbf{RQ$_4$.} \emph{What are the perceived challenges for incorporating sustainability into software engineering teams?}
    \end{rqbox}
\end{center} 

This question aims to investigate the perceived challenges, and mindset and attitudes of the participants leading to this perception.

\begin{center}
    \begin{rqbox}	    
       \textbf{RQ$_5$.} \emph{Who is responsible for making and monitoring sustainability changes in FinServCo?}
    \end{rqbox}
\end{center} 

This question aims to explore who is responsible for the implementation and also monitoring of sustainability changes.

\begin{table}[ht]
\centering
\setlength \tabcolsep{1.3 pt}
\small
\begin{tabular}{lcll}
\hline
\textbf{Role} & \textbf{Gender} & \textbf{Time Span*} & \textbf{Participation} \\
\hline
Director        & M & 4 months  & Interview (Higher Management)  \\
Director        & M & 5 years   & Interview (Higher Management)  \\
Vice President  & M & 3 years   & Interview (Higher Management)  \\
Vice President  & F & 3 years   & Interview (Higher Management)  \\
Vice President  & M & 2 years   & Interview (Higher Management)  \\
Vice President  & M & 2 months  & Interview (Higher Management)  \\
Team Lead       & M & 4.5 years & Interview (Team-Level)         \\
Team Lead       & M & 3 years   & Interview (Team-Level)         \\
Team Lead       & M & 5 years   & Interview (Team-Level)         \\
Team Lead       & M & 4.5 years & Focus Group (Team-Level)                    \\
Developer       & M & 2 years   & Interview (Team-Level)         \\
Developer       & F & 1 year    & Interview (Team-Level)         \\
Developer       & F & 3 years   & Interview (Team-Level)         \\
Developer       & M & 2 years   & Interview (Team-Level)         \\
Developer       & F & 1.5 years & Interview (Team-Level)         \\
Developer       & F & 1.3 years & Interview (Team-Level)         \\
Developer       & M & 3 years   & Focus Group (Team-Level)                    \\
Developer       & M & 4 years   & Focus Group (Team-Level)                   \\
Developer       & M & 2.7 years & Focus Group (Team-Level)                   \\
Grad Intern     & F & 1 year    & Interview (Team-Level)         \\
Grad Intern     & M & 6 months  & Interview (Team-Level)         \\
Grad Intern     & M & 1 year    & Focus Group (Team-Level)                   \\
\hline
\multicolumn{4}{r}{(*)\textbf{Time Span:} \emph{Number of years in FinServCo}}\\
\newline
\end{tabular}
\caption{Participants' Demographics} 
\label{tab:participant-demographics}
\end{table}

\subsection{Study Design}
To address our research questions, we adopted a qualitative approach using semi-structured interviews and a semi-structured focus group. The focus group was designed to facilitate group discussions and elicit shared perspectives, while the interviews were intended to uncover individual insights without the influence of colleagues or the risk of withholding information. This design was chosen to capture both individual perspectives and group dynamics, allowing us to gather rich, in-depth data on the experiences and insights of participants from FinServCo. \smallskip

\noindent {\textbf{Sampling and Saturation.}} 
We employed a multi-stage purposeful sampling method \cite{sampling} to ensure diversity among participants. In the first stage, convenience sampling was used to recruit participants who registered interest in participating in an interview or a focus group through a link shared by our gatekeeper at FinServCo. In the second stage, we used snowball sampling (recruiting new participants by asking existing participants to suggest new contacts) and maximum variation sampling (where key characteristics are identified and participants are picked with varying characteristics) to ensure diversity in terms of participants' job roles, departments, gender, and length of service with the organisation. We set our stopping criteria as code saturation \cite{saturation} (when the codebook begins to stabilise and no new sub-themes are identified in several transcripts).\smallskip

\noindent {\textbf{Interview Topics.}} We designed the initial question topics to address the research objectives. \textit{``Participant demographics''} were included to capture the professional background and daily responsibilities of participants, providing necessary data to help us contextualise their perspectives. Questions about \textit{``day-to-day challenges''} and \textit{``awareness of sustainability concepts''} were chosen to identify personal and organisational barriers to sustainability and to explore participants' familiarity with relevant frameworks, such as the United Nations Sustainable Development Goals (UN SDGs). Exploring the \textit{``meaning of sustainability''} ensured gaining a nuanced understanding of how participants conceptualised these ideas both generally and in the software engineering context. Topics on \textit{``organisational sustainability efforts,''}, \textit{``current practices,''} and their \textit{``organisation's prioritisation''} enabled us to gauge both sustainability existing initiatives and areas needing improvement. Finally, discussions on \textit{``benefits,''} \textit{``challenges,''} and \textit{``stakeholders''} provided insights into practical and strategic elements crucial for integrating sustainability into software engineering practices. Findings were mapped back to the research questions after thematic analysis, allowing the answers to our research questions to emerge from the full set of data rather than being pulled from answers to specific questions. \smallskip

\noindent {\textbf{Pilot Study and Ethics.}} 
We conducted a small pilot study of four interviews (with two members of higher management and two developers) in a different organisation to ensure our proposed data collection process and questions were appropriate. The results from these interviews are not included in the final data set, and these participants did not take part in the main study. The pilot interviews were a useful sensitising exercise, but did not result in any changes being made to the research protocol. The final semi-structured questions are included in our study material \cite{b35}. Once the design was complete, we applied for and received ethics permission to conduct the study, following the University's process.

\subsection{Data Collection}
We collected data by conducting semi-structured interviews with 11 software developers and 6 higher management participants, along with one semi-structured focus group with 5 developers. Table~\ref{tab:participant-demographics} shows the participants' demographics. The majority of the participants preferred interviews, mainly due to a lack of common availability in a hybrid working environment. Moreover, during the first focus group, we noticed that one participant dominated the discussion, whereas during interviews, participants expressed their thoughts more openly. As a result, we decided to conduct interviews for the remainder of the study.

After gaining ethics permission, our gatekeeper at FinServCo forwarded an information sheet along with a participant registration form link to employees across the organisation. All data collection was conducted either in person at the FinServCo campus or online during working hours, based on participants' availability. At the beginning of each interview and the focus group, we presented the participants with the information and consent sheet and gave them around 5 minutes to study and sign the sheet to confirm that they still wanted to participate. After collecting signatures, the voice recording was started with each participant's consent, and we took 3-5 minutes to explain the nature of the study. Then questions were asked one by one. The interviews lasted for 45-60 minutes, and the focus group lasted for 75 minutes.

The question topics were adapted based on the emerging results to keep them relevant to the organisation and to capture better information from the participants. For example, a new question topic was added about whether there was a need for a dedicated team to support SSE practices after multiple participants mentioned this in the earlier interviews. For each interview/focus group, the questions were often slightly different for each topic based on the flow of responses from the participants; hence, our study material \cite{b35} only includes the broad interview topics covered rather than the specific questions asked.

After each recorded session, we used Whisper's Large-v3 open-source transformer model by OpenAI \cite{b2} to transcribe the voice recordings. After automatic transcription, we manually checked the transcribed text to correct grammatical errors and to anonymise the transcripts by removing any personal information. Once the anonymised transcripts were ready, we deleted the voice recordings along with the file location and uninstalled Whisper.


\subsection{Data Analysis}
For data analysis we followed Braun and Clarke's inductive thematic analysis approach \cite{b3}, with slight modifications, using the NVivo software tool to manage the data. The first author conducted the detailed thematic analysis. We then used peer-debriefing \cite{peer_debriefing} as a quality check, during which the first author presented progress to the other two authors, who asked clarifying questions and discussed the chosen codes and themes.

The coding process started by reading each transcript to understand it and then rereading it to record all the key information as codes. These codes were maintained in a codebook where each row contains a code name and coded text \cite{b35}. We followed the latent approach to naming codes, where codes are formed through identifying the underlying meaning, rather than considering data at face value. In addition, the codes were initially named as long sentences to eliminate the need for description. This process was used for half of the transcription files.

After half of the transcripts were coded, some low-level codes were grouped into higher-level codes resulting in a four tier structure. Then, the rest of the transcripts were analysed and coded into this structure with new codes being created as necessary. Once all the transcripts were coded, some low-level codes became broad enough to be classified as high-level codes, and all the duplicates codes were either merged or removed where necessary. As a result, the codebook was re-organised into a three-level structure, comprising themes, sub-themes, and codes. All authors discussed and confirmed or modified these themes. The re-organisation process was iterated until all three authors agreed that the themes were stable.

To identify group-specific perspectives, we used transcript file identifiers to distinguish between higher management and developer participants during analysis. This enabled us to examine how each group related to the shared themes while maintaining a consistent coding structure.

\section{The Case Company}

Our case is a multinational, large-scale, hierarchically structured financial services organisation, FinServCo, which encompasses higher-level managers (policy makers), middle managers (policy enforcers) and software developers (policy practitioners). FinServCo employs over 80,000 people worldwide, including 5,000 at the site where we carried out this research. Staff at the site provide SE services for the company's banking operations as well as for other global FinTechs. The SE workforce is organised into divisions handling data services, technology, enterprise applications, security management, risk assessments, and more.

FinServCo uses a mixed software development approach, where some teams have recently adopted Agile methodologies, while others continue to use the Waterfall model. This distinction reflects differences in project requirements, regulatory constraints, and the complexity of legacy systems. For instance, projects that demand constant auditing or strict compliance often rely on the Waterfall model, as its sequential nature supports comprehensive documentation. In contrast, customer-facing projects tend to be developed using Agile methods to allow for iterative development, frequent feedback, and faster deployment cycles. These differences also influence how documentation is managed, potentially adding more complexity to the legacy systems. Moreover, the process of gathering requirements varies significantly as Agile allows for ongoing refinement and accepts requirement changes even mid-cycle, whereas Waterfall requires a clear and complete understanding of requirements at the start of the cycle. The organisation is actively encouraging more teams to transition to Agile by providing access to micro-courses and workshops.

When the study started in June 2024, FinServCo already had policies and frameworks in place related to achieving global sustainability. These included goals around climate action, such as aiming for net-zero emissions, investing in climate-aware startups, and supporting climate tech innovation. The company was also active in helping communities, for example, by funding sports programs, offering digital skills training to young people, and creating job pathways for service personnel and veterans. Some of the other ongoing efforts included clothing and food donation drives, sourcing food commodities used in offices in an ethical way, and offering employees subsidised interest plans to support the purchase of electric vehicles.

However, none of the sustainability initiatives had yet extended into the company’s SE teams and practices. At this point, they became interested in bringing sustainability into SE practices, and several senior managers partnered with the research team to initiate this study.

\section{Findings}
The six themes identified from the analysis represent FinServCo employees' perspectives on SSE, the challenges they face in implementing sustainable practices, and potential trade-offs between business goals and sustainability efforts. Table~\ref{tab:codebook} shows the themes and associated sub-themes. The full codebook containing themes, sub-themes and codes is available online in the supplementary material \cite{b35}.

\begin{table}[ht]
\centering
\resizebox{\linewidth}{!}{\begin{tabular}{p{0.4\linewidth}|p{0.6\linewidth}}
\hline

\textbf{Theme} & \textbf{Sub-Theme} \\
\hline

\multirow{2}{1\linewidth}{Multiple perspectives of sustainability}           & Balance of environmental-economic-social dimensions   \\
\cline{2-2}
                                                                             & Environmental - Technical concerns   \\
\cline{2-2}
                                                                             & Political strategy   \\
\cline{2-2}
                                                                             & Social dimension   \\
\cline{2-2}
                                                                             & Survive longer   \\
\cline{2-2}
                                                                             & Environmental-Social-Technical dimensions   \\
\cline{2-2}
                                                                             & Data availability  \\
\hline
\multirow{2}{1\linewidth}{Motivators to incorporate sustainability}          & Improved dependency team communications   \\
\cline{2-2}
                                                                             & Extra workload   \\
\cline{2-2}
                                                                             & Organisational benefits   \\
\cline{2-2}
                                                                             & Data availability   \\
\cline{2-2}
                                                                             & Streamlined processes   \\
\hline                                                                             
\multirow{1}{1\linewidth}{Potential trade-offs}                              & Departmental profits vs developer workload   \\
\cline{2-2}
                                                                             & Organisational profits vs employee satisfaction   \\
\cline{2-2}
                                                                             & Risk control (E.g., security vs sustainability)    \\
\cline{2-2}
                                                                             & Deadline vs sustainability   \\
\cline{2-2}
                                                                             & Efficient code vs readability   \\
\cline{2-2}
                                                                             & Functionality vs sustainability   \\
\cline{2-2}
                                                                             & Technical debt vs sustainability   \\
\hline
\multirow{2}{1\linewidth}{Challenges in integrating SSE practices}          & Clients don’t care   \\
\cline{2-2}
                                                                             & Mindset   \\
\cline{2-2}
                                                                             & Could add additional costs   \\
\cline{2-2}
                                                                             & Legacy systems   \\
\cline{2-2}
                                                                             & Not considered as a priority   \\
\cline{2-2}
                                                                             & Disturbance to work life balance   \\
\cline{2-2}
                                                                             & Poor leadership   \\
\cline{2-2}
                                                                             & Unawareness of sustainability   \\
\cline{2-2}
                                                                             & Time pressure   \\
\hline
\multirow{1}{1\linewidth}{Dedicated sustainability team}                     & Delegated ownership as a benefit   \\
\cline{2-2}
                                                                             & Parallel with security team    \\
\cline{2-2}
                                                                             & Knowledgeable leadership   \\
\cline{2-2}
                                                                             & Less workload for the middle managers   \\
\cline{2-2}
                                                                             & Difficulties to put into practice   \\
\cline{2-2}
                                                                             & May not drive large workforce   \\
\cline{2-2}
                                                                             & Poor leadership can add new challenges    \\
\hline
\multirow{2}{1\linewidth}{Migration to the cloud as a SSE solution}  & Good data redundancy techniques   \\
\cline{2-2}
                                                                             & Optimised infrastructure   \\
\cline{2-2}
                                                                             & Low employee training costs and wider talent pool   \\
\cline{2-2}
                                                                             & Pushes developers to be sustainable   \\
\cline{2-2}
                                                                             & Helps in developer personal growth   \\
\hline
\end{tabular}}
\caption{Codebook with Themes and Sub-Themes} 
\label{tab:codebook}
\end{table}

\subsection{Sustainability Perceptions} 
Participants' perceptions of sustainability referenced environmental, social, economic, technical, and individual aspects.

Higher management primarily viewed sustainability through a technical and economic lens, focusing on cloud migration as a means of operational and financial optimisation. One participant shared, \textit{``Moving from on-prem[ises servers] to CLOUD\_PROVIDER is a kind of sustainability step we are taking ... it's, like, on-demand, we're not always having resources on and utilising it.''} Data availability was another management priority, seen as central to business continuity, as one participant said, \textit{``Your data's crucial for the bank... to sustain your data and make that... readily available for everyone.''} These responses reflect a perspective of sustainability tied to system resilience and efficiency, common in the financial services context.

In contrast, developers associated sustainability with human factors such as workload and well-being. As one explained, \textit{``In a very busy environment... sometimes the workload's not sustainable.''} Their interpretations focused less on infrastructure and more on creating work environments that reduce pressure and support long-term team performance.

Environmental protection was mentioned by all roles as an important aspect of sustainability, and was understood to include practices such as energy-efficient coding, cloud computing, and reducing hardware waste. As one participant put it, \textit{``Trying to avoid destroying our planet... We can do things in a much [more] optimised way so we don't consume resources that we don't need.''}

\subsection{Motivators for Sustainability}
Management motivations centred on keeping pace with industry trends. One explained, \textit{``A lot of banks are moving their data to cloud solutions... So we are keeping up with competitors.''} Sustainability was framed as a strategic move aligned with industry innovation rather than a stand-alone ethical responsibility.

Developers, however, highlighted practical challenges as drivers for sustainability, including communication gaps and workflow inefficiencies. One participant explained, \textit{``There's no clarity... when we can start (a project)... sometimes the colleagues are busy, they don't respond on time... we have to chase a lot.''} These issues revealed a strong link between sustainability and process improvement, particularly around team collaboration.

Another motivator was the desire to shift from reactive to proactive workflows. Manual deployment processes and lack of automation were common pain points. As one participant said, \textit{``COMPANY is really bad for their deployment practices and how unsustainable they are. you will have people that end up working a lot of extra hours to get things deployed during like a release window because there's so many manual processes... that's not sustainable at all.''}

No-one mentioned environmental concerns as a motivator for introducing SSE practices.

\subsection{Potential Trade-offs}
Participants identified trade-offs between short-term goals and long-term sustainability at both organisational and team levels. While sustainable systems may offer future cost savings, initial investment was often viewed as a barrier by decision-makers.

At the team level, developers struggled to balance code quality with delivery speed. One participant noted, \textit{``There is always that balance between getting the job done quickly versus doing it right in terms of sustainability and future maintenance.''}. Furthermore, the organisation might not be willing to trade off fast pace of software development to meet release deadlines for meeting sustainability requirements: Beyond technical compromises, participants expressed scepticism about the sincerity of organisational sustainability efforts. One remarked, \textit{``It's PR at the end of the day... you're not going to advertise you're one of the biggest investors in drilling for oil... you're going to say you're investing in clean energy.''} Another added, \textit{``There are a lot of large organisations that do some sort of greenwashing... planting a tree over here and don’t look anywhere else.''} These concerns reflect how employee trust and satisfaction can be at odds with surface-level sustainability communications.

\subsection{Challenges in Integrating SSE Practices}
Cultural and structural challenges made the adoption of SSE difficult. A key issue was pressure to meet managerial expectations, even when this could increase stress as there was a lack of capacity. One participant mentioned, \textit{``People are more concerned with their manager’s opinion... if a director says this has to be done, I’m not sure how many would say no.''}

There was also a general lack of awareness and prioritisation of sustainability within teams. Some viewed it as a low-impact initiative, with unclear benefits. Security concerns, constant requirement changes, and time pressure further complicated implementation. One participant admitted, \textit{``I'm like aware of a buzzword sustainability... I don't have much knowledge of sustainability in software engineering.''}

Legacy systems and low client interest were recurring external barriers. One said \textit{``Most of my life is actually spent reading rather than coding... there's so much legacy stuff... security and risk that you always need to take into account.''}. Another added, \textit{``A lot of the clients... don’t care that much unless they’ve got their own sustainability goals.''} These factors compounded internal challenges, making SSE difficult to adopt.

\subsection{Dedicated Sustainability Team}
While there was no dedicated sustainability team within software engineering at FinServCo, participants speculated on the potential benefits and drawbacks of creating such a team, and how its presence could impact their work.

Participants believed that having a dedicated sustainability team could bring significant benefits. It could provide leadership and direction for introducing sustainable development processes, offering technical expertise and efficient solutions. One participant noted, \textit{``obviously people won't be used to developing sustainably initially. It's not necessarily a natural thing. So if you're going to set out those guidance and those policies, you should have some people that people can come and ask questions to ... So we already have that for security, right? We've got the central security team.''} Developers felt that such a team could help distribute workload more effectively by delegating ownership of sustainability initiatives, thus reducing pressure on individual engineers and middle managers as one participant explicitly said, \textit{``If you said to me along with my job, you're supposed to look out for ways in which we can be more environmentally friendly in our software engineering, that's never going to happen because it's, you know, it's just, it needs to be somebody that would need to get really deep into it''}

Despite these positive views, participants also expressed concerns that the presence of a sustainability team might lead to added pressure on SE teams, who could feel burdened by additional expectations. They also noted the risk of this dedicated team becoming isolated, and unable to drive large-scale change across different technologies and projects. Some feared that sustainability could remain an individual-level initiative rather than being embedded into organisational culture as they perceived there were no direct risks associated with not being sustainable.

\subsection{Migration to the Cloud}
Many participants highlighted the benefits of migrating to the cloud, including automated data redundancy policies, flexible infrastructure, and a reduced need for on-site hardware maintenance. These benefits could lower overall environmental impact by translating into reduced physical infrastructure, lower energy consumption, and enhanced scalability. One participant noted, \textit{``it doesn't really matter what cloud platform you use. But, again, not only is it more easy for us to access and distribute the data, but it makes you, obviously, more sustainable as well. Because, like I say, we don't now own the infrastructure. We own the data.''}

However, some participants also expressed concerns that while the cloud could help sustainability, it could also negatively impact the developer mindset as it provides easy access to powerful resources that developers might overuse without considering their environmental impact. Nonetheless, most agreed that cloud computing represents a critical tool for aiding SSE. Participants mentioned that knowledge of cloud technology is a common competency among new developers, which can help reduce training costs from the organisational perspective.

\section{Discussions}
This section discusses the novel findings of the exploratory qualitative case study we conducted at FinServCo and their implications concerning integrating SSE into finServCo's working practices.
\subsection{Higher Management Perceptions (RQ1)}

\noindent
\newline
\colorbox{bgcolor}{%
    \parbox{\linewidth}{
    \textbf{Key Finding A.1:} Higher management perceives sustainability from technical and economic perspectives, especially through cloud migration and data availability.
    }%
}

\noindent
\newline
While previous studies interpret technical and environmental SSE as system longevity \cite{b7, b9} or actions during SDLC phases \cite{b5, b8} or reducing energy consumption \cite{reducing_energy_consumption, awakening_awareness, b32}, FinServCo’s interpretation is context-specific and based on its business needs. Cloud migration is highlighted not for its environmental benefits but for its business advantages, such as data redundancy techniques, scalability, and reduced maintenance. This frames sustainability as a cost-saving measure rather than an environmental goal, which risks fostering a delegation mindset, in which sustainability responsibilities are outsourced to third-party vendors instead of being integrated into internal processes \cite{critical_cloud}. This is similar to the net-zero or carbon-neutral race in which delegation is used to achieve goals faster. While the cloud can be an efficient solution and a step towards the future, its sustainability is not guaranteed. Numerous factors need to be considered when assessing its sustainability credentials, such as choice of provider, user location, network costs, and use of virtual machines. In this case, these had not been considered.

While business continuity is an aspect of sustainability \cite{b4}, FinServCo uniquely frames sustainability around data availability because this is an operational priority in the financial services industry. This shows how sector-specific needs shape sustainability perceptions. Cloud infrastructure is perceived as a key enabler due to its built-in redundancy techniques. However, this introduces conflicts such as the potential for increased energy consumption \cite{cloud_impact}.  Also, ensuring high data availability often requires storing sensitive financial data on third-party servers, raising concerns around data security and compliance. This tension highlights the need for sustainability strategies in regulated industries to be carefully aligned with sector-specific risk frameworks, particularly where data confidentiality is of great importance.

\subsection{Software Practitioners' Perceptions (RQ2)}

\noindent
\newline
\colorbox{bgcolor}{%
    \parbox{\linewidth}{
    \textbf{Key Finding B.1:} Developers interpret sustainability mainly from individual and technical perspectives, focusing on developer workload, system efficiency and environmental factors.
    }%
}

\noindent
\newline
Developers often connect sustainability to manageable workloads and system performance \cite{b4, b5}, which extends the individual dimension of sustainability by emphasising that technical practices must support human well-being. This finding also highlights the interdependence between sustainable development processes and sustainable work environments. But software development practices at FinServCo add strain to the developer workload. Given that many teams here still follow a traditional waterfall model, the introduction of some agile practices, such as iterative planning, sustainable pacing, and retrospective adjustments, could offer a pragmatic way to improve workload balance and support individual sustainability \cite{sustaining_agile}.

\noindent
\newline
\colorbox{bgcolor}{%
    \parbox{\linewidth}{
    \textbf{Key Finding B.2:} Many developers are sceptical of FinServCo’s sustainability initiatives, viewing them as driven by PR.
    }%
}

\noindent
\newline
This is a novel insight not widely discussed in previous SSE literature. It introduces the concept of internal greenwashing where employees perceive organisational sustainability narratives as disingenuous. This scepticism can undermine employee engagement and hinder initiative success. Trust in sustainability initiatives can only be fostered through visible, meaningful, and inclusive actions that demonstrate genuine commitment. Without this, employee may lack the interest to take part, reducing the effectiveness of sustainability programs and reinforcing cynicism or passive resistance toward future initiatives.

\subsection{Perceived Needs (RQ3)}

\noindent
\newline
\colorbox{bgcolor}{%
    \parbox{\linewidth}{
    \textbf{Key Finding C.1:} Developers identify workload management, dependency team communication, and documentation as key sustainability enablers.
    }%
}

\noindent
\newline
While workload concerns are addressed very broadly from the individual perspective \cite{b4, b5}, the organisational hierarchy and process dependencies at FinServCo introduce unique frictions. Developers experience barriers not from a lack of vision, but from operational inefficiencies, such as, waiting on dependency team approvals (both on-site and on-campus), poor handovers, and repeated rework due to insufficient documentation. These bottlenecks increase stress as delivery deadlines approaching. This suggests that sustainability initiatives must go beyond high-level policy changes and address process-level frictions that impact developer productivity and well-being. Tailored interventions that improve cross-team workflows and reduce bureaucratic delays can play a critical role in advancing sustainability from the ground up.

\noindent
\newline
\colorbox{bgcolor}{%
    \parbox{\linewidth}{
    \textbf{Key Finding C.2:} There is a contrast between management and developers: Management prioritises infrastructure, while developers emphasise human-centric sustainability needs.
    }%
}

\noindent
\newline
This is a novel finding. While previous studies mention variations in perspectives of the participants across various roles \cite{b7, what_can_software_engineers_do, sustaining_agile, practitioners_prespective_gse, b8, challenges_to_incorporate_sustainability, awakening_awareness, b9, how_could_we_have_known, research_software, b33}, the divergence in sustainability priorities between organisational tiers has not been previously explored in depth. While management often frames sustainability in terms of economic gains and market competitiveness, developers’ sustainability perceptions stem from the practical challenges they encounter. This motivational asymmetry must be acknowledged when designing sustainability strategies. Otherwise, solutions may fail to resonate with or support those expected to implement them. Bridging this divergence requires cross-role internal group discussions and co-designed interventions that align both strategic objectives and lived experiences that are domain specific.

\subsection{Perceived Challenges to Incorporating SSE into Teams (RQ4)}

\noindent
\newline
\colorbox{bgcolor}{%
    \parbox{\linewidth}{
    \textbf{Key Finding D.1:} Developers mentioned low awareness, lack of training, and missing metrics as key internal challenges.
    }%
}

\noindent
\newline
Reinforcing prior research \cite{b7, what_can_software_engineers_do, sustaining_agile, practitioners_prespective_gse, b8, challenges_to_incorporate_sustainability, awakening_awareness, b9, how_could_we_have_known, research_software, b33} this finding reflects that SSE remains conceptually unclear in practice. Despite sustained attention from SSE researchers and education models \cite{everything_interrelated, customer_driven_courses, b4} over the years, the continued lack of awareness among developers reinforces a persistent gap between academic discourse and industry practice. Even interested developers lack the tools, metrics, or guidance to embed sustainability in their workflows. This highlights an urgent need for structured educational efforts within organisations, including in-house training programs, awareness workshops, and easily accessible sustainability documentation \cite{b8}. Without deliberate investment in knowledge-building, sustainability risks remaining an abstract ideal rather than a practical component of day-to-day software engineering work.

\noindent
\newline
\colorbox{bgcolor}{%
    \parbox{\linewidth}{
    \textbf{Key Finding D.2:} FinServCo’s performance-driven culture is a major cultural obstacle due to misaligned incentives and the fear of change.
    }%
}

\noindent
\newline
This is a novel contribution. While resistance to change is a known challenge \cite{b8, challenges_to_incorporate_sustainability, challenges_that_challenge}, this study additionally uncovers how performance cultures, where employees aim to impress and managers race for promotions, actively de-prioritise sustainability. This reveals how organisational values and reward systems directly influence whether sustainability is seen as viable or burdensome. In such an environment, sustainability is often de-prioritised in favour of short-term visibility and career advancement. Additionally, concerns about the additional time and extra complexity of implementing sustainability changes further reinforce resistance. Without systemic shifts in how success is measured, recognised, and rewarded, sustainability is likely to remain a peripheral concern. This highlights the need for long-term, organisation-wide cultural change that embeds sustainability into performance metrics and professional development frameworks.

\noindent
\newline
\colorbox{bgcolor}{%
    \parbox{\linewidth}{
    \textbf{Key Finding D.3:} Legacy systems and lack of client pressure are seen as external and structural barriers.
    }%
}

\noindent
\newline
While the burden of legacy systems \cite{sustaining_software_ecosystems, b33} and disinterest of client during requirement engineering \cite{b8} were previously mentioned, this study highlights organisational and external constraints, such as the high cost of sustainability training, resistance to altering legacy systems due to fear of risk, and a lack of client interest at any stage of development pose significant barriers to sustainability adoption reducing its strategic value. These factors create a risk-averse environment where sustainability is perceived as an optional or burdensome investment, especially when not directly tied to revenue or regulatory compliance. In financial services industries, the absence of client-driven pressure can lead to sustainability being de-prioritised in strategic decision-making. However, rather than waiting for external demand, organisations like FinServCo could take a more proactive role by embedding sustainability into their offerings and creating incentives for clients to engage with more sustainable practices. This would position the company not only as a compliant actor, but as a leader in driving sector-wide change.

\subsection{Responsibility and Monitoring of SSE (RQ5)}

\noindent
\newline
\colorbox{bgcolor}{%
    \parbox{\linewidth}{
    \textbf{Key Finding E.1:} Many participants recommended a centralised dedicated sustainability team, but others advocated for shared responsibility.
    }%
}

\noindent
\newline
This is a novel finding. While literature discusses sustainability ownership abstractly (from either top-down, bottom-up, or mixed approach) \cite{advocate, b5, b33, b4}, this study surfaces a concrete governance tension. A centralised team offers structure and expertise but risks creating a culture of delegation, where sustainability responsibility is seen as someone else's job. Distributed ownership encourages embedded responsibility but can dilute accountability. For sustainability efforts to be effective, they must be embedded at both organisational and individual levels. A hybrid approach that combines centralised leadership with team-level sustainability champions can help balance expertise and engagement, and drive the sustainability change more effectively. In this model, the central team facilitates, monitors, and supports efforts, while distributed champions encourages SE teams and ensure that sustainability becomes part of everyday decision-making and team culture.

\noindent
\newline
\colorbox{bgcolor}{%
    \parbox{\linewidth}{
    \textbf{Key Finding E.2:} Participants drew parallels between the potential sustainability team and existing structures such as the security team.
    }%
}

\noindent
\newline
Drawing direct analogies between sustainability governance and established models like security teams represents a creative, practice-informed contribution not widely explored in existing research. While drawing inspiration from security governance provides a practical blueprint for SSE integration, the organisational context differs in important ways, and it is unlikely to succeed. Unlike security, which carries immediate financial, legal and reputational risks, sustainability lacks the same urgency at FinServCo. This limits top-down motivation at this stage, where long-term benefits of SSE are still unclear. A more effective approach could involve appointing sustainability champions within teams to initiate local awareness and action, while a central sustainability team is developed in parallel. This central team can emerge through multi-role working groups that surface contextual risks, shared needs, and opportunities, ultimately creating an informed, balanced, and organisation-wide governance structure.

\section{Threats to Validity}

Robson \cite{b19} highlights three primary threats to validity in qualitative research: \textbf{description validity}, \textbf{interpretation validity}, and \textbf{theory validity}. Each of these threats, if unaddressed, can introduce biases that impact the credibility of a study. Below, we address how these threats were handled in this study and we discuss generalisation of the results from this study.

\textit{Description validity} refers to the accuracy and completeness of the data as recorded during data collection. If the raw data (such as interviews or observations) is inaccurately captured or incomplete, it poses a significant threat to the study’s credibility. In this study, all interviews and focus groups were recorded throughout and later fully transcribed, ensuring the complete capture of participants' responses. This eliminates any risk of selective or incomplete data recording, thereby reducing the potential for description validity threats.

\textit{Interpretation validity} concerns the accuracy with which the researcher interprets the data. This threat arises when researchers impose their own understanding, expectations, or biases onto the participants' responses. To address this threat, the coding of data was conducted without any pre-existing theoretical lens, allowing themes to be identified directly from the data. After one author formed initial themes, two other authors reviewed the results to ensure the accuracy and objectivity of the interpretations through peer debriefing. This collaborative review process ensured that personal biases were minimised, and interpretations were cross-validated by multiple researchers.

\textit{Theory validity} involves consideration of multiple theories naturally coming from the data. This threat arises when researchers stick to the first theory formed during the analysis and ignore any subsequent theories which may arise. This study doesn't suffer from theory validity as the main objective is to understand different perspectives rather than develop a single theory.

We also employed triangulation techniques: \textbf{observer triangulation} (more than one observer was involved in the focus group to ensure that data was not misinterpreted) and \textbf{data triangulation} (more than one data collection strategy was used to capture a wide range of responses) to handle these specific biases. As mentioned by Robson \cite{b19}, triangulation can bring some contradictions. However, for our exploratory study, data triangulation approach provided further insights.

\textit{Generalisability.} We conducted this study to investigate the specific SSE needs of one organisation, FinServCo, to facilitate the integration of sustainability into its SE practices. Case studies are not generalisable by their very nature because they are context-specific \cite{empirical_standards_casestudy}. There is a gap in the literature concerning case studies that investigate an organisation's context-specific needs, and our study addresses this research gap. One can find ways to put SSE into practice only by understanding an organisation's context-specific needs. Our study can be an example for organisations as a prerequisite for an action research study to identify their SSE-related problems and provide solutions for integrating them into their workflow. We anticipate that other companies across the financial services industry and other industries can find many findings from this study relevant, but this will need to be confirmed by further research.

\textit{Further Limitations.} While we reached code saturation among the set of employees we accessed, given that FinServCo has a very large employee force, we could have missed individuals who could have provided other perspectives. We tried to recruit participants from as diverse characteristics as possible to mitigate this limitation.

\section{Conclusion and Future Work}

According to the Brundtland report \cite{b11}, sustainability, by its very nature, is a concept involving trade-offs as ideal sustainability across all dimensions is not possible. Instead, priorities must be negotiated depending on context, goals, and constraints. Our study reflects this reality within a financial services organisation, where we found that different roles perceive and prioritise sustainability differently. Higher management emphasises economic goals such as infrastructure optimisation and business continuity, while developers focus on individual and human-centric needs like workload balance and well-being. These varying perspectives highlight the lack of a shared understanding within the organisation. Discussions, a key part of incorporating sustainability, were also lacking in the organisation. Recognising this, our next step is to conduct a workshop to facilitate open dialogue between roles, aiming to co-create a shared, context-specific approach to SSE that can guide actionable change.

After establishing this shared understanding of sustainability, the next step will be to engage in action research. This will focus on designing and implementing targeted SSE interventions that align with identified priorities. By doing so, we can provide deeper insights into the practical challenges and opportunities associated with integrating SSE practices in industry practice. We hope this process will uncover success stories and effective strategies that can serve as models and frameworks for other industries.

While many of our participants see cloud computing as having potential for improving sustainability, others raised concerns about its environmental impact and possible overuse of resources. Moreover, organisational bottlenecks such as legacy systems, ineffective team communication, and lack of ownership could slow progress. This highlights the lack of industry-accepted metrics for quantifying sustainability efforts. Hence, another crucial step will be developing concrete metrics for SSE practices. These should assess sustainability from various perspectives, considering the trade-offs between them.

While many participants expressed an interest in having a dedicated sustainability team, further exploration is needed to understand how such a team could effectively integrate across different departments and projects without adding new blockers or delays to current workflows. Sustainability champions at the team and department levels could also be helpful. Incentives for sustainable efforts could be a catalyst to bring in this change.

Given that this is a single case study of one FinServ organisation, it would be beneficial to undertake further case studies within this industry. Understanding how different financial institutions approach sustainability in software engineering could reveal best practices and common obstacles, providing actionable insights for the industry. Similar studies could also be replicated across other industries to understand both general and context-specific sustainability needs and challenges.

\section*{Acknowledgment}
We would like to thank all the interview and focus group participants for taking part in this study.

\bibliographystyle{ieeetr}
\bibliography{citations}

\begin{thebibliography}{10}

\bibitem{b11}
W.~C. on~Environment and Development, ``Our {Common} {Future},'' 1987.

\bibitem{b12}
C.~Becker, R.~Chitchyan, L.~Duboc, S.~Easterbrook, B.~Penzenstadler, N.~Seyff, and C.~C. Venters, ``Sustainability {Design} and {Software}: {The} {Karlskrona} {Manifesto},'' in {\em 2015 {IEEE}/{ACM} 37th {IEEE} {International} {Conference} on {Software} {Engineering}}, vol.~2, pp.~467--476, May 2015.
\newblock ISSN: 1558-1225.

\bibitem{b21}
B.~Purvis, Y.~Mao, and D.~Robinson, ``Three pillars of sustainability: in search of conceptual origins,'' {\em Sustainability Science}, vol.~14, pp.~681--695, May 2019.

\bibitem{case_study_runeson}
P.~Runeson and M.~H\"{o}st, ``Guidelines for conducting and reporting case study research in software engineering,'' {\em Empirical Softw. Engg.}, vol.~14, p.~131–164, Apr. 2009.

\bibitem{b35}
A.~Anonymous, ``Supplementary data for perspectives, needs and challenges for sustainable software engineering teams: A finserv case study,'' Apr. 2025.

\bibitem{b6}
S.~Betz, D.~Lammert, and J.~Porras, ``Software {Engineers} in {Transition}: {Self}-{Role} {Attribution} and {Awareness} for {Sustainability},'' 2022.

\bibitem{b4}
R.~Heldal, N.-T. Nguyen, A.~Moreira, P.~Lago, L.~Duboc, S.~Betz, V.~C. Coroamă, B.~Penzenstadler, J.~Porras, R.~Capilla, I.~Brooks, S.~Oyedeji, and C.~C. Venters, ``Sustainability competencies and skills in software engineering: {An} industry perspective,'' {\em Journal of Systems and Software}, vol.~211, p.~111978, May 2024.

\bibitem{b7}
I.~Groher and R.~Weinreich, ``An {Interview} {Study} on {Sustainability} {Concerns} in {Software} {Development} {Projects},'' in {\em 2017 43rd {Euromicro} {Conference} on {Software} {Engineering} and {Advanced} {Applications} ({SEAA})}, pp.~350--358, Aug. 2017.

\bibitem{b9}
M.~Souza, R.~Haines, and C.~Jay, {\em Defining {Sustainability} through {Developers}’ {Eyes}: {Recommendations} from an {Interview} {Study}}.
\newblock Jan. 2014.

\bibitem{b8}
R.~Chitchyan, C.~Becker, S.~Betz, L.~Duboc, B.~Penzenstadler, N.~Seyff, and C.~C. Venters, ``Sustainability design in requirements engineering: state of practice,'' in {\em Proceedings of the 38th {International} {Conference} on {Software} {Engineering} {Companion}}, {ICSE} '16, (New York, NY, USA), pp.~533--542, Association for Computing Machinery, May 2016.

\bibitem{b5}
M.~Levy, E.~C. Groen, K.~Taveter, D.~Amyot, E.~Yu, L.~Liu, I.~Richardson, M.~Spichkova, A.~Jussli, and S.~Mosser, ``Sustaining human health: {A} requirements engineering perspective,'' {\em Journal of Systems and Software}, vol.~204, p.~111792, Oct. 2023.

\bibitem{b33}
L.~Karita, B.~C. Mourão, and I.~Machado, ``Towards a common understanding of sustainable software development,'' in {\em Proceedings of the {XXXVI} {Brazilian} {Symposium} on {Software} {Engineering}}, {SBES} '22, (New York, NY, USA), pp.~269--278, Association for Computing Machinery, Oct. 2022.

\bibitem{b32}
S.~R.~A. Ibrahim, J.~Yahaya, H.~Salehudin, and A.~Deraman, ``The {Development} of {Green} {Software} {Process} {Model},'' {\em International Journal of Advanced Computer Science and Applications (IJACSA)}, vol.~12, no.~8, 2021.
\newblock Number: 8 Publisher: The Science and Information (SAI) Organization Limited.

\bibitem{sampling}
A.~J. Onwuegbuzie and N.~L. Leech, ``A {Call} for {Qualitative} {Power} {Analyses},'' {\em Quality \& Quantity}, vol.~41, pp.~105--121, Feb. 2007.

\bibitem{saturation}
M.~M. Hennink, B.~N. Kaiser, and V.~C. Marconi, ``Code {Saturation} {Versus} {Meaning} {Saturation}: {How} {Many} {Interviews} {Are} {Enough}?,'' {\em Qualitative Health Research}, vol.~27, pp.~591--608, Mar. 2017.

\bibitem{b2}
A.~Radford, J.~W. Kim, T.~Xu, G.~Brockman, C.~McLeavey, and I.~Sutskever, ``Robust {Speech} {Recognition} via {Large}-{Scale} {Weak} {Supervision},'' Dec. 2022.
\newblock arXiv:2212.04356.

\bibitem{b3}
V.~Braun and V.~Clarke, ``Thematic {Analysis},'' Oct. 2021.

\bibitem{peer_debriefing}
J.~W. Creswell and D.~L. Miller, ``Determining validity in qualitative inquiry,'' {\em Theory Into Practice}, vol.~39, no.~3, pp.~124--130, 2000.

\bibitem{reducing_energy_consumption}
Z.~Ournani, R.~Rouvoy, P.~Rust, and J.~Penhoat, ``On {Reducing} the {Energy} {Consumption} of {Software}: {From} {Hurdles} to {Requirements},'' in {\em Proceedings of the 14th {ACM} / {IEEE} {International} {Symposium} on {Empirical} {Software} {Engineering} and {Measurement} ({ESEM})}, {ESEM} '20, (New York, NY, USA), pp.~1--12, Association for Computing Machinery, Oct. 2020.

\bibitem{awakening_awareness}
E.~Jagroep, J.~Broekman, J.~M.~E. van~der Werf, P.~Lago, S.~Brinkkemper, L.~Blom, and R.~van Vliet, ``Awakening {Awareness} on {Energy} {Consumption} in {Software} {Engineering},'' in {\em 2017 {IEEE}/{ACM} 39th {International} {Conference} on {Software} {Engineering}: {Software} {Engineering} in {Society} {Track} ({ICSE}-{SEIS})}, pp.~76--85, May 2017.

\bibitem{critical_cloud}
``A {Critical} {Review} of {Cloud} {Computing}: {Researching} {Desires} and {Realities}.''

\bibitem{cloud_impact}
F.~Khomh and S.~A. Abtahizadeh, ``Understanding the impact of cloud patterns on performance and energy consumption,'' {\em Journal of Systems and Software}, vol.~141, pp.~151--170, July 2018.

\bibitem{sustaining_agile}
L.~Barroca, P.~Gregory, K.~Kuusinen, H.~Sharp, and R.~AlQaisi, ``Sustaining {Agile} {Beyond} {Adoption},'' in {\em 2018 44th {Euromicro} {Conference} on {Software} {Engineering} and {Advanced} {Applications} ({SEAA})}, (Prague), pp.~22--25, IEEE, Aug. 2018.

\bibitem{what_can_software_engineers_do}
R.~Chitchyan, J.~Noppen, and I.~Groher, ``What {Can} {Software} {Engineering} {Do} for {Sustainability}: {Case} of {Software} {Product} {Lines},'' in {\em 2015 {IEEE}/{ACM} 5th {International} {Workshop} on {Product} {Line} {Approaches} in {Software} {Engineering}}, pp.~11--14, May 2015.

\bibitem{practitioners_prespective_gse}
I.~Manotas, C.~Bird, R.~Zhang, D.~Shepherd, C.~Jaspan, C.~Sadowski, L.~Pollock, and J.~Clause, ``An empirical study of practitioners' perspectives on green software engineering,'' in {\em Proceedings of the 38th {International} {Conference} on {Software} {Engineering}}, (Austin Texas), pp.~237--248, ACM, May 2016.

\bibitem{challenges_to_incorporate_sustainability}
M.~Bamiduro, I.~Fatima, P.~Lago, and S.~Vos, ``Challenges in {Incorporating} {Sustainability} {Practices} in the {Software} {Lifecycle},'' in {\em Software {Business}} (E.~Papatheocharous, S.~Farshidi, S.~Jansen, and S.~Hyrynsalmi, eds.), (Cham), pp.~284--290, Springer Nature Switzerland, 2025.

\bibitem{how_could_we_have_known}
J.~Porras, C.~C. Venters, B.~Penzenstadler, L.~Duboc, S.~Betz, N.~Seyff, S.~Heshmatisafa, and S.~Oyedeji, ``How {Could} {We} {Have} {Known}? {Anticipating} {Sustainability} {Effects} of a {Software} {Product},'' in {\em Software {Business}} (X.~Wang, A.~Martini, A.~Nguyen-Duc, and V.~Stray, eds.), (Cham), pp.~10--17, Springer International Publishing, 2021.

\bibitem{research_software}
M.~Rosado~de Souza, R.~Haines, M.~Vigo, and C.~Jay, ``What {Makes} {Research} {Software} {Sustainable}? {An} {Interview} {Study} with {Research} {Software} {Engineers},'' in {\em 2019 {IEEE}/{ACM} 12th {International} {Workshop} on {Cooperative} and {Human} {Aspects} of {Software} {Engineering} ({CHASE})}, pp.~135--138, May 2019.
\newblock ISSN: 2574-1837.

\bibitem{everything_interrelated}
B.~Penzenstadler, S.~Betz, C.~C. Venters, R.~Chitchyan, J.~Porras, N.~Seyff, L.~Duboc, and C.~Becker, ``Everything is {INTERRELATED}: {Teaching} {Software} {Engineering} for {Sustainability},'' in {\em 2018 {IEEE}/{ACM} 40th {International} {Conference} on {Software} {Engineering}: {Software} {Engineering} {Education} and {Training} ({ICSE}-{SEET})}, pp.~153--162, May 2018.

\bibitem{customer_driven_courses}
O.~Cico, L.~Jaccheri, and A.~N. Duc, ``Software {Sustainability} in {Customer}-{Driven} {Courses},'' in {\em 2021 {IEEE}/{ACM} {International} {Workshop} on {Body} of {Knowledge} for {Software} {Sustainability} ({BoKSS})}, pp.~15--22, June 2021.

\bibitem{challenges_that_challenge}
P.~Gregory, L.~Barroca, H.~Sharp, A.~Deshpande, and K.~Taylor, ``The challenges that challenge: {Engaging} with agile practitioners’ concerns,'' {\em Information and Software Technology}, vol.~77, pp.~92--104, Sept. 2016.

\bibitem{sustaining_software_ecosystems}
Y.~Dittrich, ``Software engineering beyond the project – {Sustaining} software ecosystems,'' {\em Information and Software Technology}, vol.~56, pp.~1436--1456, Nov. 2014.

\bibitem{advocate}
B.~Penzenstadler, H.~Femmer, and D.~Richardson, ``Who is the advocate? {Stakeholders} for sustainability,'' in {\em 2013 2nd {International} {Workshop} on {Green} and {Sustainable} {Software} ({GREENS})}, pp.~70--77, May 2013.

\bibitem{b19}
C.~Robson, ``Real {World} {Research}, 3rd {Edition} {\textbar} {Wiley}.''

\bibitem{empirical_standards_casestudy}
P.~Ralph, N.~bin Ali, S.~Baltes, D.~Bianculli, J.~Diaz, Y.~Dittrich, N.~Ernst, M.~Felderer, R.~Feldt, A.~Filieri, B.~B.~N. de~França, C.~A. Furia, G.~Gay, N.~Gold, D.~Graziotin, P.~He, R.~Hoda, N.~Juristo, B.~Kitchenham, V.~Lenarduzzi, J.~Martínez, J.~Melegati, D.~Mendez, T.~Menzies, J.~Molleri, D.~Pfahl, R.~Robbes, D.~Russo, N.~Saarimäki, F.~Sarro, D.~Taibi, J.~Siegmund, D.~Spinellis, M.~Staron, K.~Stol, M.-A. Storey, D.~Taibi, D.~Tamburri, M.~Torchiano, C.~Treude, B.~Turhan, X.~Wang, and S.~Vegas, ``Empirical standards for software engineering research,'' 2021.

\end{thebibliography}

\end{document}